# Notebook articles: towards a transformative publishing experience in nonlinear science

*Authors*: Cristel Chandre (corresponding author)
    cristel.chandre@univ-amu.fr - ORCID: 0000-0003-3667-259X
  Jonathan Dubois
    jdubois@pks.mpg.de - ORCID: 0000-0002-6976-5488

## Introduction

    Our scientific dissemination strategy is centered on one *fundamental unit / paradigm*: the scientific/research article. It is where research results on a specific question are reported and disseminated in a broad scientific community. A set of scientific articles defines the corpus of knowledge in a given field - here nonlinear science - from which scientific communication is derived to the benefit of society at large, from textbooks to popular science.
Ideally, a research article should be clearly written, should be accurate and correct in its conclusions, and should rely on a transparent methodology. It is often a small visible part of months to years of collaborative research with a complex path leading to the reported, often summarized, results. Once peer-reviewed and validated by the community, nowadays these articles are published as a single and easily readable document, most often a PDF file. They are then broadly accessible on the website of the publisher/journal, and each article is uniquely identified with a direct link, e.g., through a digital object identifier. Sometimes, this file can be accompanied on the side by data, graphs, videos or supplemental material containing additional information supporting the results presented in the article. As researchers, we collectively recognize the instrumental role played by research articles, as a trusted link between the labs and the society at large from policy makers to the general public, to the point where publishing articles defines the core of our work as researchers.

> *"Scientific methods evolve now at the speed of software (...). And yet the basic means of communicating scientific results hasn't changed for 400 years."*
> [Somers, 2018]

    However, changes in our dissemination practices are needed: The reported results are often insufficiently or partially substantiated in the article, e.g., they are presented for a selected

set of parameters for practical reasons, and they are often difficult and challenging to reproduce, even for the close community. Usually, the authors make several compromises in writing a research article: These result from a subtle balance between the completeness of the reported study and a clearly conveyed message, mostly due to the finite size of the article, the finite attention span of the reader and the goal to get the manuscript accepted for publication. In particular, details are not always shown for the benefit of an efficient display of the results. Unfruitful attempts are often not reported, since they do not shine a good light on the reported results, potentially impacting acceptance in prominent journals and future citations. In addition, the parameters of the study or simulations are often chosen so as to highlight the reported results in a clearer and more comprehensive way and to better support the conclusions. Nonetheless, alternative attempts might constitute useful and helpful pieces of evidence to the community, to better and fully grasp the results. This shortcoming is closely linked to the positive-results bias, which is an increasingly recognized weakness of the current dissemination paradigm.

Furthermore, the results reported in articles are rarely straightforward to reproducible, and correctness is often difficult to assess, e.g., when data are not accessible or codes not provided. The reviewers and the readers rely on their preconceived ideas of the field to judge/evaluate the correctness of the presented results. This, of course, introduces a bias in favor of the prevailing ideas in a given community (which leads to the so-called confirmation bias). As a consequence, we believe there is an urgency to fully embrace reproducibility in scientific publishing, to remedy all these shortcomings and biases [Munafò, 2017]. On a short time scale, this can be addressed by adding data and codes as supplemental material. On a longer time scale, attempts to remedy these shortcomings invite us to reinspect and challenge our central model, the fundamental unit of the scientific dissemination, the research article.

Some research communities have advocated and put into practice the use of lab notebooks as a primary record of research and as a way to legally support the conclusions presented in research articles. These lab notebooks are contemporaneous notes on the research as it unfolds, allowing the community to follow, verify and reproduce the various steps leading to the conclusions summarized in the articles, including both successful or unsuccessful attempts. To a large extent, these lab notebooks represent a very valuable source of scientific information, but they are often not accessible. Even for the close community they are often not easy to read, as they follow the complex path of the research work as it unfolds. With the migration of research dissemination towards digital media, some of these notebooks moved to electronic lab notebooks. This step is convenient since it allows a larger research community to adopt this healthy work practice, and since it has a clear potential to contribute to the Open Science initiative. However, the current implementation of lab notebooks has inherent shortcomings: Currently there is no unified system to read, store and share them, contrary to what has been set up for research articles. For example, an electronic lab notebook as a side document stored on a personal computer is often very volatile, and its lifetime is short even if stored on an institutional repository. In addition, the link with the research article is not often made clear, and the notebooks add another layer to the jungle of dissemination tools without being clearly findable, accessible and interoperable. Nevertheless, these notebook documents have the potential to provide useful results to the community if made available, e.g., as it was done for the discovery of gravitational waves [Abbott, 2016]. In this context, the Open-Notebook Science initiative [Bradley, 2007] advocates the central role of notebooks as a publicly-available primary record of research. In our opinion, there is not enough emphasis on notebooks in scientific dissemination, especially in nonlinear science.

Here we propose to reexamine the paradigm of research dissemination by discussing and advocating another fundamental unit, the "notebook article". To be clear, our purpose is not to discuss how to enhance the reader's experience by adding content and functionality, or by enriching their browsing/reading experience, to the standard research article. Our goal is to discuss, define and implement another paradigm tailored to the needs of a scientific community, by taking full advantage of cloud computing solutions.
In this viewpoint, we describe what the basic requirements of a fundamental unit of scientific dissemination should be, addressing open science, reproducibility and Findable, Accessible, Interoperable, Reusable (FAIR) data principles [Wilkinson, 2016]. We then describe some related attempts in the recent past and in different fields. We propose a format for a notebook article in nonlinear science which is designed for the practitioners, then we propose a technical solution. We provide an example of what such a notebook article could look like in comparison with its standard version.

## Requirements

Typical articles in nonlinear science are based on analytical calculations, mathematical derivations, numerical simulations or manipulation/representation of data obtained from experimental measurements or simulations. What makes nonlinear science a community is mostly based on shared interests and shared tools and methods. From this observation, we establish a list of requirements that shape the format of the proposed fundamental unit of dissemination we advocate, and which we refer to as a "notebook article" in what follows.

- First and foremost, the notebook article should be a *single document*, uniquely and perennially identified with a digital object identifier. It must contain all the features of a standard research article which are widely accepted as best practices in scientific dissemination. Notebook articles must be correct, well-documented and complete. They should convey clear and interesting messages to the community. In addition, they should be robust, perennial and easily accessible.
- Notebook articles should contain/embed all the elements to check the validity of their reported conclusions, and more precisely to reproduce the results presented in the notebook article, e.g., figures and quantitative estimates. This notion of *reproducibility* is linked to Open Science initiatives as it allows a broader community to have access to the core of the research work in a transparent way [Masuzzo, 2017; Lasser, 2020]. The notebook articles should be readily executable with the use of cloud computing platforms. The possibility to run the notebook article on cloud computing platforms

- should be embedded in the notebook in a transparent way, without any requirements from the reader to have specific knowledge on cloud computing or even on coding.
- Notebook articles should be easily *accessible*, readable/downloadable on all standard platforms, devices and operating systems, without the need for third-party softwares except for a web browser. It should be possible to read and run (in the sense of verify and reuse) these articles regardless of the wealth of an institution or country.
- Notebook articles should be *easy to read*. All readers should be able to automatically generate a printable document (e.g., a PDF file) with different levels of reading from a single notebook article: for instance, a short version containing the essential elements of the proposed research, and a more detailed version for interested readers and practitioners. This hierarchical reading contributes to the goal of publishing less and publishing better.
- With running the document on a cloud computing platform to check and reproduce the results, readers should be able to *interactively* use notebook articles as research tools. For instance, the conclusions of an article could be dependent on some parameters: Varying some parameters could be of interest to the reviewers to check the validity and robustness of the conclusions reported in the article, to better assess the depth of the conclusions, and to formulate a better informed recommendation. This will contribute to protecting research results against cognitive biases [Munafò, 2017]. It will also be a valuable asset to other readers and members of the community, as the codes and data could be reused to address other questions without the need to redo the codes, e.g., as it was done in neuroscience with the Neurolibre repository of notebooks [Neurolibre]. It contributes to saving time and valuable resources.
- Data management: The raw data used for the production of the results should obey the FAIR principles. In particular, these raw data should be available to the community on a general-purpose or on a domain-specific repository, and should be connected with the notebook article in question, whether it is fully embedded or clearly tethered to the notebook. In any case, the codes leading to any modifications of the raw data should be embedded in the notebook article. We refer to [Wilkinson, 2016] for an in-depth discussion on data management and its FAIR guiding principles.
- We also mention the possibility to modify the notebook article in a continuous way, with a version/git system. If granted by the publisher/editors, authors would have the possibility to update the content of the notebook article after it is published, for instance by complementing their study with incremental additions/modifications which are useful for the community. Each of these modifications would be dated and moderated. The typical example we have in mind is the continuous update of a wikipedia article. It is an appealing concept which might contribute to publishing less and publishing better, even though we realize that, at this stage, this possibility raises more questions than it solves problems as far as scientific dissemination is concerned (e.g., related to moderating these changes, modifications in the contributors, or dating specific discoveries). So we will not dwell on this requirement.

The previous requirements address two observations made in [Gavish, 2011]: "Text-based publications are not enough" and "Every result is detached from its creation process". With the proposed requirements, which push the publication beyond text-based, we argue that the creation process is reflected in the presentation of the results in the notebook article to the extent adequately judged by the authors. In summary, a notebook article should be a single document, uniquely and persistently identified, accessible, and easy to read, run and interactively re-use. It should embed all the elements for research reproducibility.

## Open science and reproducible research

Reproducibility ought to be an essential pillar of scientific production, and considered as a keystone of scientific methodology. It helps the community to check the validity of the results and conclusions reported in an article, and it contributes to transfer knowledge in an open-science framework. Usually, reproducibility is indirectly addressed by a relevant and detailed description of the methods used, whether it is in the article itself or presented as supplemental material. More rarely, practical reproducibility is addressed by rendering additional files, such as codes and data, available on personal websites or on the cloud. Only experienced practitioners are then able to run these files on their computers to check the validity, as they need to have the necessary knowledge, the technical infrastructure, and the softwares to compile and run these files. We argue that this does not constitute an optimal way to practically implement reproducibility since, even if the files are accessible, the technical gaps might still constitute a serious obstacle to practical reproducibility. In addition, different readers will most likely generate different outputs as the environment (e.g., operating system, softwares, computer specifications) is usually attached to the reader and not to the file itself (for instance, see the issues in reproducibility in computational fluid dynamics [Mesnard, 2017]). Ideally, all readers should have the exact same experience in generating the results as the authors had in reaching their conclusions. As a consequence, we argue that all the elements making the research reproducible should be naturally *embedded* in notebook articles.

Code and data, whether embedded or tethered, should be part of the notebook article itself whenever feasible. Running these codes should not be uniquely restricted to experienced readers nor to experienced programmers, but should be accessible to the broader nonlinear science community as research tools. This practice also contributes to the objective of developing research resources for the community by encouraging safe reuse through appropriate licensing. In addition, reproducibility has to be ensured on long timescales, not just at the time of submission for publication. We do not think that hosting the resources on institutional websites can ensure sustainability, since this is not their primary mission. Nowadays, powerful technical solutions exist to ensure this paramount requirement, relying heavily on cloud computing solutions. The time is ripe for fully integrating cloud computing tools, not only into our daily work as researchers, but also into our dissemination strategy. We believe that integrating these tools will shape our work as researchers in similar ways it is shaping our daily lives as citizens. Cloud computing solutions have the potential to shape our dissemination strategy to allow a full and practical embedding of reproducibility in the way we publish results. The improved resilience of cloud computing solutions ensures that this reproducibility has the potential to be sustainable. We understand that notebook articles do not constitute a flawless solution, but we believe it is a significant step forward compared to the standard-article format. We refer to [Akhlaghi, 2020] for a more in-depth discussion of the potential problems and some possible solutions.

## Article of the future

There have been many attempts to rethink what should or will be the article of the future. Many of these attempts are within specific scientific communities, e.g., climate science, neuroscience, geoscience, metabolomics to name a few. One of the main and early driving forces behind these changes is computer science. Some early attempts along the directions mentioned above date back to more than a decade ago. For instance, Elsevier's Executable Papers Grand Challenge (2011) in computer science [Elsevier, 2011] stated that "*In an 'executable paper', authors have the ability to embed chunks of executable code and data into their papers, and readers may execute that code within the framework of the research article. The executable paper combines the narrative of a traditional scholarly paper with embedded computational experiments that run chunks of code on prespecified or interactively provided datasets, producing verifiable results*." The article of the future follows in part the principles of literate programming [Knuth, 1984], which is intimately linked to reproducibility.
Even if, at the time of the challenge, the technical solutions were not mature enough to trigger changes in our dissemination strategy, it is remarkable to note that some elements of the proposed solutions we detail below were already hinted at and discussed in the winning papers of this challenge [Nowakowski, 2011; Van Gorp, 2011; Gavish, 2011].

More recently, in Geosciences, a very interesting initiative was launched to define the "geoscience paper of the future" [Gil, 2016a, 2016b] including considerations of the best practices in software and data sharing. The forum leading this initiative recognized the importance of reproducible publications and executable papers, and the sharing of data and softwares through third-party repositories (GitHub, etc…). We advocate for similar solutions for the nonlinear science community, as detailed below.
In our opinion, the name of the concept "paper of the future" gives it a science-fiction flavor, which is nowadays less and less appropriate, since technical solutions are now ready to be rather easily implemented and are within reach. To the best of our knowledge, the closest initiative addressing the above-mentioned requirements is the Image Processing On Line journal [IPOL], which is an interesting but isolated initiative not followed up by major publishing houses. We also mention the promising publishing venue Open Research Europe intended for Horizon 2020 and Horizon Europe beneficiaries [Open Research Europe, 2020].

## Proposed format for notebook articles

The format we advocate for nonlinear-science scholarly articles is based on widely used notebooks such as Mathematica notebooks, Matlab live scripts and Jupyter notebooks. For unfamiliar readers, a notebook is generally a sequence of cells of 3 different types: (*i*) *markdown cells* contain formatted text with LaTeX equations describing the proposed research, (*ii*) *code cells* contain the codes used to support the conclusions of the proposed research and, when executed, the results are reported in (*iii*) *output cells*. Using an appropriate combination of these cells, one can easily devise a format that contains the main markers of a scholarly article, but at the same time, can be interactive and executable. We note that such a format was also described in [Nowakowski, 2011] with "static content", "execution input data" and "interactive result visualization".

Some rules on how to structure and write a notebook are provided in [Rule, 2019]. We stress some of these important rules such as "document the process, not just the results" and "use cell divisions to make steps clear", which provide not only a guideline for reproducible research in an open science context, but also suggest a method and a blueprint to achieve it.

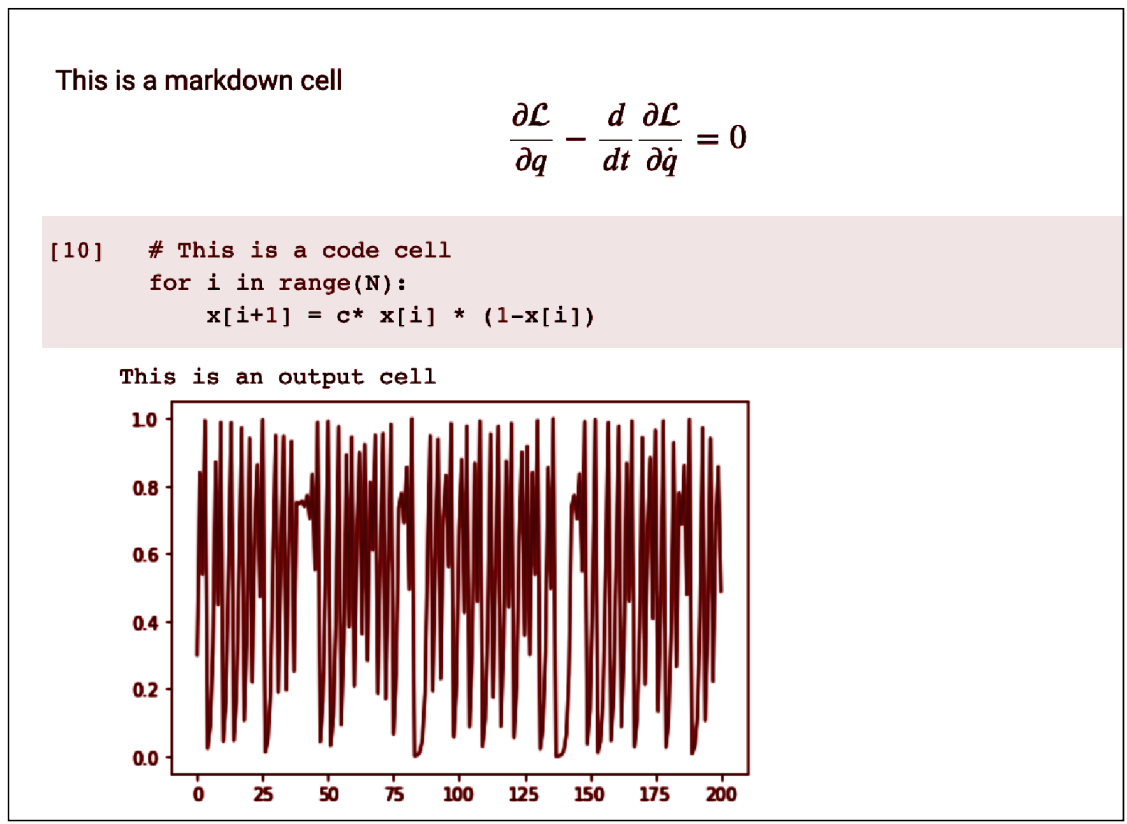

*Figure 1: A notebook typically contains three types of cells: markdown cells, code cells and output cells.*

At first glance, notebook articles will look exactly the same as standard articles with a title, list of authors, affiliations, abstract, introduction, sections with figures, conclusions, acknowledgments and references. The main difference is that data and codes are embedded or tethered in the notebook, to compose an autonomous executable and interactive unit. For instance, any readers (even non-experts in coding) are able to re-run all figures, or run them using different parameters than the ones selected by the authors for producing the original figures and for reaching their conclusions. Any reader can open the code cells used for producing figures, as a way to get deeper insights into the methodology, e.g., the selected algorithms, devised computational methods and visualization tools.

A notebook offers the possibility to include interactive input forms through user-friendly widgets in order to change the parameters of the computation, allowing the reader to test the conclusions of the article for different values of the parameters than the ones used in the article itself, in a very interactive way. These features do not require any programming skills from the reader's side.

For each figure, the environment could contain each of the 3 cells: (i) a markdown cell, with caption and explanatory text, (ii) a code cell, with parameters clearly visible as widgets or form fields for interactivity purposes, and (iii) an output cell, generated by the computing environment, which contains the graphical display of the figure.

Notebooks offer additional advantages to tailor the display of articles to the needs of the reader: For instance, the authors could tag some markdown cells as in-depth reading which would be hidden when the article is first opened. This could be a way to embed supplemental materials for interested readers or to provide more in-depth reading without resorting to additional documents, and without cluttering the layout of the article at first sight. By expanding some of the cells of the notebook, the reader would get access to in-depth content and a different view of the article. The format we advocate exploits the versatility of notebooks. It provides authors with opportunities to structure the layout of their notebook article in an innovative way, to allow for a better dissemination of their research results.

The technical solution for the implementation of the above-described requirements is the key to demonstrating the feasibility and power of notebook articles in research dissemination. It is only in very recent years that notebook articles can be envisaged, not just as articles of the future but as articles of tomorrow.

The main element for advocating drastic changes in our dissemination practices is the increasing availability and pervasiveness of cloud computing. To a large extent, cloud computing platforms implicitly invite us to rethink the concept of our dissemination plan, and they already suggest a possible framework with the combination of three elements: a document format, a repository and an interactive and reproducible environment.

As a crucial element of the technical solution, we believe it is essential that all the elements of the proposed framework are based on open-source solutions and softwares. This ensures a perennial accessibility of the articles without relying on the existence and maintenance of a specific software by third-party companies.

In particular, we advocate that a promising format for scientific publishing is the Jupyter notebook format (.ipynb document) [Jupyter; Kluyver, 2016; Mendez, 2019; Perkel, 2018; Randles, 2017]: Jupyter notebooks are web-based computational documents composed of an ordered sequence of cells containing formatted text (including LaTeX equations), codes (which could be written in variety of programming languages) and figures. These notebooks could be stored on a Github repository hosted by the publisher on the journal's website, as is currently done for research articles. For instance, we propose the use of BinderHub [Jupyter, 2018] handled by the publisher through cloud computing platforms. It allows the execution of the deposited notebook using the same computing environment, e.g., regarding the specific softwares to run the codes and their dependencies. For a tutorial on the combination Jupyter/GitHub/BinderHub, we refer to [Mendez, 2019] and references therein.

The technical solutions we advocate are readily available on major cloud computing platforms such as Google Cloud Platform through Google Colaboratory [Colab]. If willing, Google or other major cloud computing actors already have the capacity to launch scientific journals of notebook article types. Note that a BinderHub [Jupyter, 2018] could use computing time bought from a cloud computing company as this is actually what is done for mybinder [mybinder]. Therefore, the proposed technical solution does not require the publisher to possess and maintain a cloud computing infrastructure themselves, but instead they could rely on the major actors of cloud computing, while focusing on their main core of business, publishing.

One of the major advantages of such cloud-embedded solutions is its wide accessibility with a simple internet connection and a web browser. It would benefit institutions and research groups from low-income countries, since this solution does not require the purchase and maintenance of expensive equipment to produce or to read, use and run notebooks. We believe that a modern vision of academic publishing needs to embrace notebook articles as this format offers a more inclusive and more transparent way of disseminating research. The journal of the future might look like or include a github repository of notebooks, hosted and maintained by the publisher. Each notebook article will contain a relevant and clear license file, specifying the conditions of use for codes and data, and a clear documentation to run and reuse the listed material. Finally, each notebook article will be uniquely identified with a digital object identifier.

## Role of the publisher/editors

One of the roles of the publisher is to act as an independent notary public, certifying when an article has been submitted and accepted, and keeping record on how and why an article has been accepted. A publisher ensures accessibility to the articles on large time scales, time-scales much larger than the typical lifetime of a research group. This is still the case in the notebook model we advocate. In this respect, data and codes as part of the article should be hosted by the publisher, not in third-party repositories, to guarantee the perennity of the access to the published research work. This access could follow traditional routes, such as subscription-based or open-access journals, but notebooks also open the way for alternative and more innovative routes: For instance, the publisher could decide to make part of the notebook accessible in open-access while restricting the access of code cells and markdown cells tagged as advanced-reader only. Likewise, the whole notebook, or the capacity to run it, could be accessible to subscribers only.
Our goal here is not to advocate for a specific publishing framework, which is beyond the scope of this viewpoint, but to illustrate the various routes offered by this very versatile article format.

We also believe that notebook articles provide opportunities for the peer-review process. Our purpose is not to offer our views on the peer review process of scientific articles itself, but to lay out what notebook articles will allow as peer-review is concerned. Like for standard articles, the rules for peer-reviewing notebook articles would be set by the publisher. They would follow already established best practices in the community and address the publishing requirements/standards set for the journal. At the pace the number of publications increases, the peer-review process will soon (and to some extent already is) a burden to the scientific community. There is a well-recognized need to experiment and define new peer-review models which allow for a close monitoring of what is considered as the corpus of knowledge in a given field and, at the same time, accompany its growth without risking asphyxiation.

Notebook articles promise to contribute significantly to the objective of publishing less and publishing better. The combination of a pre-publication peer review and a post-publication one is an example of how the versatile format of a notebook article can contribute to enriching the discussion on peer-review procedures in scientific publishing. A pre-publication peer-review is a necessary step to select the highest quality articles possible. We advocate that it should be based on the short PDF extracted from the document to reduce the burden on reviewers. A post-publication peer-review will assess the article in greater length, in particular addressing reproducibility and robustness of the conclusions in the long term. This post-publication peer review could be open to readers, performed by the targeted community, in the form of a forum hosted by the publisher and moderated by its editors on the website of the notebook article.

## An example of a "notebook article": double ionization of atoms in intense laser fields

Writing notebooks is now very common in science both for teaching and for research collaborations. This model is less common in some areas, especially regarding research dissemination. Here we provide an example of a "notebook article", i.e., a notebook meant for scientific publication and dissemination, which is the strict analog of the research article [Mauger, 2009]. In this article, the classical motion of two electrons inside atoms driven by a strong laser pulse is investigated by theoretical and numerical methods. Two routes for the double ionization of the atoms are identified and a scenario for the escape of the two electrons as a function of the laser parameters, in particular its intensity, is established. The corresponding notebook article is a Jupyter notebook (.ipynb format) publicly available on GitHub at [github.com/cchandre/Notebook-Article](github.com/cchandre/Notebook-Article).

In order to experience all the benefits of the notebook in comparison with the standard article, a link/badge to Google Colaboratory to read, modify, and run the notebook is included on the GitHub page of this notebook (this can also be done using mybinder.org -- the badge is also provided on GitHub). Note that all the elements of the original article [Mauger, 2009] are in this notebook when it is first opened. Moreover, the figures of [Mauger, 2009] can be reproduced without being an expert in simulations, here the coding is done in Python. The cell 'Parameters of the notebook', which defines the packages used, the classes and the basic functions, should be first executed after setting the main parameters of the model (by clicking on the play icon next to the cell). Then each figure can be executed independently (by clicking on the play icons right next to the "Execute Figure" label) after setting the parameters associated with the figure in question. As a consequence, we argue that reproducibility of the research summarized in [Mauger, 2009] is fully embedded in the corresponding notebook article. Its parameters can be changed, and new figures can be produced to check the robustness of the conclusions drawn in [Mauger, 2009]. This interactivity serves as an independent way to check the results, and it can readily be used to address similar questions in the close scientific community.

## Potential issues

The proposed notebook solution is not devoid of potential issues which will need to be discussed and resolved before a broad initiative can be launched. These issues include the recognition of publishing notebook articles in scientific careers, intellectual property issues in a very broad sense and the conservatisms naturally associated with any change of paradigm. The changes in dissemination paradigms, notably involving notebooks, have to be accompanied by a discussion on research evaluation, not based on quantitative indicators, but more on the content of the scientific production (see also [DORA, 2012]). Other issues include backwards compatibility of the published codes, and technical debt for the solution to be deployed for notebooks.

The stakeholders - from the researchers to the publishers - might express concerns in coping with this dissemination tool: For the authors, a notebook will be too long to prepare, e.g., related to commenting and documenting codes and data, and it will require the also time-consuming acquisition of new knowledge with a very uncertain reward. This issue is made acute as the research competition is fierce in some communities. For the publishers, adopting this new format for their journals means an increasing amount of work, in particular, to check the technical validity of the submitted and then published notebooks.

We believe that the discussions and exchanges of best practices around these issues should be able to lift potential obstacles related to conservatism. We argue that notebooks represent a more in-depth contribution to the dissemination effort, a richer content to be shared, and as such they will likely be more visible.

In growing research communities, we also argue the need to publish less and publish better. This will contribute to a more efficient transfer of knowledge to the general public. The notebook articles partly address this need by offering a structured/hierarchical reading with potentially different levels of reading, e.g., first reading and in-depth reading as described above.

We will not dwell upon issues related to intellectual property (see [Stodden, 2019] for more details): these issues should be addressed within a specific community, and they will strongly depend on the types of codes and data shared in a notebook. In [Barnes, 2010], Barnes discussed issues related to releasing the codes, and provided compelling arguments in favor of publishing the codes, whether these are well written and well commented or not. We are aware of the potential feeling of losing a competitive advantage by publishing source codes or data, potentially impacting the careers of the researchers involved. We think that the practice of publishing notebooks inside a given community will convince the hesitating researchers of the relevance of this publishing channel by showcasing a real added value to their dissemination strategy and overall to their research.

In the end, the authors should be responsible for the way they disseminate their results, tailored to the proposed research. Authors already ask themselves these kinds of questions when selecting a journal, e.g., broad audience versus specialized journal, and an article type, e.g., letter, research article, review or comment. We advocate that adding notebooks as an article type in existing journals, or in journals of a new kind, will add a meaningful new avenue of disseminating results that will more closely match the proposed research. It is a way for the authors to potentially devise a dissemination strategy which fully integrates and embraces the versatile and powerful tools afforded by cloud computing platforms.

Other important stakeholders, notably research institutions and funding agencies, play an increasing role in the way research is disseminated. For instance, we mention the Open access and Data management initiatives of the European Union's H2020 framework program for research and innovation, and European Union's strategy for Open Science [EU]. The versatile format of the notebook will naturally allow researchers to accommodate the needs of their institutions and their funding agencies, for instance, regarding open access and data management, without compromising the interests of other stakeholders.

## Conclusions and outlook

Notebooks are increasingly used in our daily life as researchers as they provide efficient ways to document individual research and to increase the efficiency of collaborations. The goal of this viewpoint is to promote this work practice to the realm of scientific publishing in nonlinear science.

Powered by cloud computing solutions, notebook articles have the potential to transform our dissemination practices. They are not meant to replace standard research articles since one format does not fit all, but they offer an alternative, efficient and more ethical way to disseminate research, by providing a practical tool allowing more interactive and more reproducible research. We believe that nonlinear science is an ideal community to implement these new ways of disseminating research results in the form of notebook articles. These articles are easily and practically interactive and they are reproducible; they go beyond the standard reading experience provided by a PDF or an HTML display of the article. Without altering the main goal of research dissemination, namely the presentation of well-grounded conclusions on specific questions, some of these notebook articles could act as research tools to the community, when properly engineered. We believe that this possibility will increase the visibility of the reported research.

Major cloud computing platforms already possess the technical infrastructure to implement the proposed solution for scientific dissemination. The main question for us is whether the stakeholders in scientific publishing among them the researchers, the publishers, the funding agencies and the research institutions are willing to embark in such a drastic change in our dissemination strategy.

With this viewpoint, we aim to interrogate and stimulate discussions in the nonlinear science community on the feasibility and support of such an initiative and, in the long term, the opportunity to experiment this strategy with a new form of journals in nonlinear science. It is technically feasible, and we believe that this is in near-sight of academic publishing.

## Acknowledgements and disclaimer

We acknowledge useful discussions with Simon A. Berman, François Mauger, Manos Mavrakis and Bradley A. Shadwick. We also acknowledge useful inputs from the editors of Communications in Nonlinear Science and Numerical Simulation. The views and opinions expressed in this article are those of the authors. They are not meant to reflect any policy of the publisher nor the opinions of the editorial board of Communications in Nonlinear Science and Numerical Simulation.

## Competing interests

CC is the Editor-in-chief of Communications in Nonlinear Science and Numerical Simulation and an expert-evaluator in the European Union's framework programmes for research and innovation.

## CRediT author statement

**C. Chandre:** Conceptualization, Writing - Original Draft, Software.
**J. Dubois:** Writing - Review & Editing, Software.

## References

- [Abbott, 2016] B.P. Abbott et al. (LIGO Scientific Collaboration and Virgo Collaboration), "Observation of Gravitational Waves from a Binary Black Hole Merger", Phys. Rev. Lett.

116: 061102 (2016); doi: 10.1103/PhysRevLett.116.061102; GitHub link: github.com/minrk/ligo-binder
- [Akhlaghi, 2020] M. Akhlaghi, R. Infante-Sainz, B.F. Roumeka, D. Valls-Gabaud, R. Baena-Gallé, "Towards long-term and archivable reproducibility", pre-print arXiv:2006.03018; see also zenodo.org/record/3911395
- [Barnes, 2010] N. Barnes, "Publish your computer code: it is good enough", Nature 467: 753 (2010); doi: 10.1038/467753a
- [Bradley, 2007] J.C. Bradley "Open Notebook Science Using Blogs and Wikis". Nature Precedings (2007); doi: 10.1038/npre.2007.39.1
- [Colab] colab.research.google.com/
- [DORA, 2012] The San Francisco Declaration on Research Assessment (DORA); sfdora.org/
- [Elsevier, 2011] Elsevier's Executable Papers Grand Challenge (2011) in computer science www.journals.elsevier.com/the-journal-of-logic-and-algebraic-programming/news/introducing-executable-papers and www.elsevier.com/about/press-releases/science-and-technology/elsevier-announces-winners-of-the-executable-paper-grand-challenge
- [EU] ec.europa.eu/research/openscience
- [Gavish, 2011] M. Gavish, D. Donoho, "A Universal Identifier for Computational Results", Procedia Computer Science 4: 637 (2011); doi: 10.1016/j.procs.2011.04.067
- [Gil, 2016a] Y. Gil (Ed.), "The Scientific Paper of the Future: OntoSoft Training", Zenodo (2016); doi: 10.5281/zenodo.159206; see also the special issue in Earth and Space Science agupubs.onlinelibrary.wiley.com/doi/toc/10.1002/(ISSN)2333-5084.GPF1
- [Gil, 2016b] Y. Gil, C.H. David, I. Demir, B.T. Essawy, R.W. Fulweiler, J.L. Goodall, L. Karlstrom, H. Lee, H.J. Mills, J.-H. Oh, S.A. Pierce, A. Pope, M.W. Tzeng, S.R. Villamizar, X. Yu, "Towards the Geoscience Paper of the Future: Best Practices for Documenting and Sharing Research from Data to Software to Provenance", Earth and Space Science 3: 388 (2016); doi: 10.1002/2015EA000136
- [IPOL] Image Processing On Line ipol.im
- [Jupyter] jupyter.org/
- [Jupyter, 2018] Jupyter *et al.*, "Binder 2.0 - Reproducible, Interactive, Sharable Environments for Science at Scale", Proceedings of the 17th Python in Science Conference (2018); doi: 10.25080/Majora-4af1f417-011
- [Kluyver, 2016] T. Kluyver, B. Ragan-Kelley, F. Pérez, B. Granger, M. Bussonnier, J. Frederic, K. Kelley, J. Hamrick, J. Grout, S. Corlay, P. Ivanov, D. Avila, S. Abdalla, C. Willing, Jupyter Development Team, "Jupyter Notebooks - a publishing format for reproducible computational workflows", Positioning and Power in Academic Publishing: Players, Agents and Agendas. F. Loizides and B. Schmidt (Eds). IOS Press (2016); doi: 10.3233/978-1-61499-649-1-87
- [Knuth, 1984] D.E. Knuth, "Literate Programming", The Computer Journal 27: 97 (1984)
- [Lasser, 2020] J. Lasser, "Creating an executable paper is a journey through Open Science", Communications Physics 3: 143 (2020); doi: 10.1038/s42005-020-00403-4

- [Masuzzo, 2017] P. Masuzzo, L. Martens, "Do you speak Open Science? Resources and tips to learn the language", PeerJ Preprints (2017); doi: 10.7287/peerj.preprints.2689v1
- [Mauger, 2009] F. Mauger, C. Chandre, T. Uzer, "Strong field double ionization: The phase space perspective", Phys. Rev. Lett. 102: 173002 (2009); doi: 10.1103/PhysRevLett.102.173002 ; GitHub link: github.com/cchandre/Notebook-Article
- [Mendez, 2019] K.M. Mendez, L. Pritchard, S.N. Reinke *et al*., "Toward collaborative open data science in metabolomics using Jupyter Notebooks and cloud computing", Metabolomics 15: 125 (2019); doi: 10.1007/s11306-019-1588-0
- [Mesnard, 2017] O. Mesnard, L.A. Barba, "Reproducible and replicable computational fluid dynamics: It's harder than you think", Computing in Science & Engineering 19: 44 (2017); doi: 10.1109/MCSE.2017.3151254
- [Munafò, 2017] M.R. Munafò, B.A. Nosek, D.V.M. Bishop, K.S. button, C.D. Chambers, N. Percie du Sert, U. Simonsohn, E.-J. Wagenmakers, J.J. Ware, J.P.A Ioannidis, "A manifesto for reproducible science", Nature Human Behaviour 1: 0021 (2017); doi: 10.1038/s41562-016-0021
- [mybinder] mybinder.org/
- [Neurolibre] Neurolibre conp-pcno.github.io/
- [Nowakowski, 2011] P. Nowakowski, E. Ciepiela, D. Harezlak, J. Kocot, M. Kasztelnik, T. Bartynski, J. Meizner, G. Dyk, M. Malawski, "The Collage Authoring Environment", Procedia Computer Science 4: 608 (2011); doi: 10.1016/j.procs.2011.04.064
- [Open Research Europe, 2020] open-research-europe.ec.europa.eu/
- [Perkel, 2018] J.M. Perkel, "Why Jupyter is data scientists' computational notebook of choice", Nature 563: 145 (2018); doi: 10.1038/d41586-018-07196-1
- [Randles, 2017] B. M. Randles, I. V. Pasquetto, M. S. Golshan, C. L. Borgman, "Using the Jupyter Notebook as a Tool for Open Science: An Empirical Study", 2017 ACM/IEEE Joint Conference on Digital Libraries, Toronto, ON (2017); doi: 10.1109/JCDL.2017.7991618
- [Rule, 2019] A. Rule, A. Birmingham, C. Zuniga, I. Altintas, S.-C. Huang, R. Knight *et al*., "Ten simple rules for writing and sharing computational analyses in Jupyter Notebooks", PLoS Comput Biol 15: e1007007 (2019); doi: 10.1371/journal.pcbi.1007007
- [Somers, 2018] J. Somers, "The Scientific Paper is Obsolete", The Atlantic (April 2018) www.theatlantic.com/science/archive/2018/04/the-scientific-paper-is-obsolete/556676/
- [Stodden, 2009] V. Stodden, "The legal framework for reproducible scientific research: Licensing and copyright", Computing in Science & Eng. 11: 35 (2009); doi: 10.1109/MCSE.2009.19
- [Van Gorp, 2011] P. Van Gorp, S. Mazanek, "SHARE: a web portal for creating and sharing executable research papers", Procedia Computer Science 4: 589 (2011); doi: 10.1016/j.procs.2011.04.062
- [Wikinson, 2016] M.D. Wilkinson, M. Dumontier, I.J. Aalbersberg, G. Appleton *et al*., "The FAIR Guiding Principles for scientific data management and stewardship", Scientific Data 3: 160018 (2016); doi: 10.1038/sdata.2016.18